\documentclass[12pt]{article}

\newcommand{\be}{\begin{equation}}
\newcommand{\ee}{\end{equation}}
\newcommand{\ba}{\begin{eqnarray}}
\newcommand{\ea}{\end{eqnarray}}

\begin{document}
\setlength{\baselineskip}{.5cm}
\renewcommand{\thefootnote}{\fnsymbol{footnote}}
\newcommand{\lp}{\left(}
\newcommand{\rp}{\right)}

\begin{center}
\centering{\bf \Large The sharp peak-flat trough pattern and 
critical speculation}
\end{center}

\vskip 1cm

\begin{center}
\centering{B. M. Roehner $^1$ and D. Sornette $^{2,3}$\\
$^1$  L.P.T.H.E. \quad University Paris 7, \\
roehner@lpthe.jussieu.fr\\
$^2$ L.P.M.C. CNRS UMR6622 and
Universit\'e de Nice-Sophia Antipolis, B.P. 71
06108 NICE Cedex 2, France\\
$^3$ IGPP and ESS department\\
UCLA, Box 951567, Los Angeles, CA 90095-1567\\
sornette@naxos.unice.fr}
\end{center}

\vskip 3cm

{\bf Abstract}

We find empirically a characteristic sharp peak-flat trough pattern
in a large set of commodity prices. We argue that the sharp peak structure
reflects an endogenous inter-market organization, and that peaks may be seen
as local ``singularities'' resulting from imitation and herding.
These findings impose a novel stringent constraint on the construction of models.

\vskip 3cm
PACS numbers\,: 01.75+m  Science and society\\
				05.40.+j Fluctuation phenomena, random processes, and Brownian motion\\
				05.70.Jk Critical point phenomena\\
				89.90.+n Other areas of general interest to physicists\\

\pagebreak

\section{Introduction}

In the stock market, returns over long period of times are often 
mainly due to rare large upward price variations that occur over
a tiny fraction of the total trading time\,: the US equity index
$S\&P500$ for instance has gained an average of 16\% a year from 
$1983$ to $1992$ and $80 \%$ of this return stems from forty days of
trading, i.e. less than $1.6 \%$ of all working days. This
property is shared by other markets and other assets. The
prices of commodities that we will investigate here is no exception.

The prices of most commodities are also characterized by rare and
sudden bursts of apparently outlying values. A typical and spectacular
example is provided by the evolution of the price
of gold and silver\,; in the half century since 1950, these prices
experienced one huge peak that lasted for two decades and
resulted in a multiplication of the (deflated) price by a factor
of the order of $10$. 

These two examples outline the importance of both rare events and
the effect of correlations in the determination of market price time series.
The gaussian paradigm of
independent normally distributed price increments \cite{Bachelier,Samuelson} has long
been known to be incorrect with many attempts to improve it. Econometric nonlinear
autoregressive models with conditional heteroskedasticity (ARCH) \cite{Engle} and
their generalizations \cite{Bollerslev} as well as jump-diffusion models
\cite{Jump} capture only imperfectly the volatility (variance) correlations and the
fat tails of the probability density distribution (pdf) of price variations. 
Alternatively, the fat tail properties of the full pdf (corresponding to a one-point
statistics) has been described by a L\'evy law \cite{Mandelbrot} for cotton
and other commodities
and more recently by a truncated L\'evy flight \cite{Stanley,Arneodo} for 
equities or 
by a superposition of Gaussian pdf's with log-normally distributed variance
\cite{Peinke}. A recent decomposition of the volatility (standard deviation) of 
return data across scales of several financial time series has revealed
the existence of a causal information 
cascade from large scales to fine scales that expresses itself in the
volatilities \cite{cascade}. 

In very liquid markets of equities and
foreign exchanges for instance, correlations of price variations 
are extremely small, as any significant correlation would lead to
an arbitrage opportunity that is rapidly exploited and thus washed out.
Indeed, the fact that there are almost no correlations between price
variations in liquid markets can be understood from the following
simple calculation presented in \cite{Bouchaud} that we slightly extend.
For the sake of illustration, let us assume that the price 
variations $\delta x$ over the
time interval $\delta t$ are gaussian correlated stationary variables of zero
average. Time is expressed as a multiple of $\delta t$. The time series
$\delta x_1, \delta x_2, ..., \delta x_t$ is denoted by the column vector 
$X_t$. The correlation function is defined by
\be
\langle \delta x_n \delta x_k\rangle \equiv C(n,k)~,
\label{corr}
\ee
which is $C= \langle X X^T \rangle$ in matricial notation.
If the correlation functions are the only
constraints, then it is straightforward to show, under the assumption of a maximum
entropy principle, that the knowledge of the correlation functions is fully embedded
in that of the following multivariable probability distribution for the variables
$\delta x_n$
\be
P(X_t) = P_0 \exp \biggl(-{1 \over 2} X_t^{T} C^{-1} X_t \biggl)~ .
\label{distribb}
\ee
where $X_t^{T}$ denotes
the transpose of $X$, i.e. the uni-row matrix whose $n$th row is $\delta x_n$.
$C^{-1}$ is the inverse of the correlation matrix.
Expression (\ref{distribb}) can be rewritten explicitely as
\be
P(X) = P_0 \exp \biggl(-{1 \over 2} \sum_{n,m+1}^t C^{-1}(n,m) X_n X_m
\biggl)~ . 
\label{didgfibb}
\ee
This expression shows that, conditioned to the past values of 
the variations $\delta x_0, ... \delta
x_{t-1}$, the probability density function of  $\delta x_t$ is then given by  
\be
P(\delta x_t) = P_0 \exp \biggl( -\frac{C^{-1}(t,t)}{2} [\delta x_t - m_t]^2
\biggl) ~,
\ee
where $P_0$ is a normalization factor. The average $m_t$ of $\delta x_t$ conditioned
to the past is 
\be
m_t \equiv \frac{\sum_{i=0}^{t-1} C^{-1}(i,t) 
\delta x_i}{C^{-1}(t,t)}~.
\ee
It may be non-zero due to the presence of correlations. 
A simple trading strategy consists in buying a unit stock if $m_t > 0$
(expected future price increase) and selling if $m_t < 0$ (expected future price
decrease). The average gain is then $\langle |m_t| \rangle > 0$. Let us consider the
short range limit where only $C^{-1}(t,t)$ and $C^{-1}(t,t-1)$ are non-zero and
$\delta t$ is then equal to the correlation time which is typically $5$ 
minutes for liquid markets. The average return over one
correlation time is then ${1 \over e} \langle {| \delta x| \over x} \rangle
\approx 3.7 ~~10^{-4}$ for $\langle {| \delta x| \over x} \rangle  \approx 10^{-3}$.
Over a day, this gives an average gain of $0.59 \%$ which accrues to $435 \%$ per
year when return is reinvested or $150 \%$ without reinvestment! Such small
correlations would lead to substantial profits if transaction costs and 
other friction phenomena like slippage did not exist {\footnote{
Slippage refers to the fact 
that market orders are not always executed at the order price due to 
limited liquidity and finite human execution time.}.
Counting the transaction costs of about $c \approx 0.1 \%$ and since 
one must on average
modify the position once in a correlation time to achieve the above performance,
this leads to an effective average gain
per transaction equal to  $e^{-1} \langle {|\delta x| \over x} \rangle - c$ which
becomes negative in our numerical application! Since
the scaling law $|\delta x| \approx ({\delta t \over T})^H$ (with $H \approx 0.6$) 
\cite{Olsen}, holds for time scales less than about a day, a
given level of transaction cost $c$ allows correlations to develop over a maximum
time $\delta t \approx T (ec)^{1 \over H}$ (such that 
$e^{-1} \langle {|\delta x| \over x} \rangle = c$)
without allowing arbitrage opportunities. The small
level of transaction costs in efficient modern markets thus explains the low level of
correlations of price variations, completing the proof of the above assertion.

The liquidity and efficiency (level of transaction costs, slippage) of markets thus
control the degree of correlation that is compatible with the almost absence of
arbitrage opportunity. In other words, if one detects the possibility to make
money in a given market, the difficulty for entering the market and the costs
limit the concrete realization of this opportunity.
Less liquid markets thus allow the appearance of stronger correlations
that may take more intricate forms. Here, 
we point out the existence of a more subtle kind of price correlations
in commodities, namely the repeated existence of 
``sharp peak-flat trough'' (SP-FT) patterns , which we will study. 
Identifying patterns in economic and financial
prices has a long history and is often refered to as ``technical analysis''.
Technical analysis in finance can be broadly defined as the 
study of financial markets, mainly using graphs of stock prices as a function
of time, in the goal of predicting future trends \cite{Elliott}. 
A lot of efforts has been 
developed both by academic and trading institutions and more
recently by physicists (using some of their statistical tools developed to deal
with complex times series) to analyse past data to get informations on 
the future. We notice that the Dow Jones US index was also characterized
by a dominance of SP-FT
from 1875 to 1935, but since 1950, the
sharp peaks have left place to smoother structures. This may be seen as
the consequence of a more efficient arbitrage progressively destroying
those patterns that would provide arbitrage opportunities.

Our motivation for this study is to provide a novel 
characterization, with as few fitting parameters as possible,
of economic time series which can be 
useful for sparse data. This is
a useful practical alternative (and/or complement) to the determination of 
statistical distributions and correlations of price changes. The problem is that
in order to obtain a probability distribution function
with reasonable precision, especially in
the range of large price changes, huge records of several thousand
prices are required. In contrast, the shape of the peaks can
be analysed on fairly small samples. 
In restricting ourselves to the structure of the
peaks, we aim at a more modest goal which is to focus
on a simple sub-structure of the whole time series. The justification
of this approach is simply that all scientific endeavors (may) succeed when
restricting their aim to simple enough problems which, when understood, can be
extended and generalized towards the more complex problems.

The main message of this paper is to point out that the shape
of major price peaks of commodities varies within rather narrow limits. More
specifically, the peaks are of what we call the sharp peak type\,:
 if $p$ denotes the price, 
$ x = \ln (p) $ turns its concavity upwards both before and
after the turning point\,; mathematically this means that the 
(suitably coarse-grained) second 
derivative of $x(t)$ with respect to 
time is positive close to the peak turning point.
When looking at a time series as a whole, this structure leads to the
appearance of patterns that can be described as SP-FT.
As a consequence, the graph of $\ln p(t)$
is {\it not} bottom-top symmetric\,: if the graph is rotated by $180^o$ or
inverted upside-down, the resulting figure will no longer resemble the graph of
a genuine commodity price series. Qualitatively, this observation has been made some
time ago \cite{Drame}. Our purpose here is to characterize it quantitatively and
propose a theoretical basis for it. It is important to realize that the 
existence of these SP-FT patterns
put strong constraints on any general theory
commodity prices. Previous attempts \cite{DeatonLaroque}
have already highlighted the difficulty of generating peaks having
realistic amplitude and frequency. If a general dynamical model for the
price behavior of commodities would be available, the pattern
of the peaks could be derived as a  simple consequence. Some
models in this spirit have indeed been investigated by a number of authors
\cite{Labys1,Labys2,Weimar},  but such a global approach turns out to be extremely
difficult. 

The shape constraint that we
describe in this paper is very demanding. Thus, in order for 
a model to qualify, not only should it be compatible with the one-point
(probability distribution) and two-points statistics (correlation functions of the 
volatilities), it should also correctly account for the 
SP-FT patterns. It is indeed important to 
constraint as much as possible the model construction as it has been
found in the past for instance that several models could account equally well for 
the probability distribution \cite{Stanley,Arneodo,Peinke,cascade}. 
We do not claim that the 
SP-FT patterns always exist for all commodities but when there
are present they should be taken into account. 

The second important message of the paper is that the sharp peak structure may
reflect an endogenous inter-market organization, and that peaks may be seen
as local ``singularities''. This conclusion extends and generalizes 
the view that the largest possible peaks in the stock market,
preceeding the large crashes in the stock market, can be modelled
as special critical crises 
\cite{crash1,crash2,crash3,crash4,crash5,crash6,crash7,crash8}\,:
the underlying hypothesis is that stock market crashes may be caused by the
slow buildup
of powerful subterranean forces that come together in one critical instant.
The use of the word ``critical'' is not purely literary here\,: in
mathematical terms, complex dynamical systems such as the stock market can go through
so-called ``critical'' points, defined as the explosion to infinity of a
normally well-behaved quantity.

\vskip 0.5cm
The paper is organized as follows. 
In section 2, we describe the SP-FT patterns for
the case of pre-twentieth century wheat markets. These markets
have the advantage of displaying a large number of price peaks. We
are thus in an ideal position for a systematic quantitative
analysis. In section 3, we consider some twentieth century
commodity markets. The same pattern will be seen to hold in a number
of important instances. In addition, we examine some speculative
price bubbles outside the sphere of commodity markets in order to
emphasize both the similarities and the discrepancies. In section 4,
we discuss some requirements to be fulfilled by a future theory 
and outline such a theory.

\section{The sharp peak-flat trough pattern in the case of wheat
markets}

\subsection{Wheat markets}

Before 1850, wheat was the key product in the economies of Western
Europe. Its production was the major task of the agricultural
sector which employed over two thirds of the total manpower; its
consumption was a crucial element in the diet of the masses; 
and finally its trade represented a large part of
international trade. Because the price of wheat was of crucial
importance both for the economy and for the public welfare, it was
carefully recorded in every state. Some of these records extend
uninterrupted from the 15th to the 20th century, and they contain
many huge peaks. 

A superficial view would ascribe such peaks solely to the
occurrence of defavourable meteorological conditions. In fact,
meteorological hazards were only the triggering factors in a 
more complex process. As will be seen subsequently, the average
duration of the price peaks is of the order of four years, a
feature that clearly speaks against the meteorological
explanation since poor weather conditions during four years in
a row are rather unlikely. As a matter of fact, the crucial role
was played by speculation and monopolistic practices as has been
convincingly documented by many historians \cite{Biollay,Modeste,Persson,Usher}. 
Indeed, hoarding was common practice, not only by traders and
retailers but also by the consumers as is shown by the following
excerpt from Biollay \cite{Biollay}\,: ``A survey conducted by the French State
Council described how people bought up the wheat market at the
first signs of a coming wave of high prices, and how they stored
large amounts of wheat for future use when prices would have
doubled or tripled''.
In short, the mechanisms by which the pre-twentieth century wheat
markets operated were basically the same as those at
work in modern commodity markets. One should notice
an important difference, namely no futures markets
existed. But some practices, such as buying the wheat in spring
(i.e. several months before the harvest), already announced the
mechanism of the forward markets. 
Because for wheat markets, there are so many long and reliable
price records, we are in an ideal position to carry out a statistical
analysis of price peaks. Such an analysis involves the following
steps\,: 
\begin{enumerate}
\item the selection of the peaks\,; 
\item the definition of the parameters by which such peaks can be described\,; 
\item the statistical analysis of those parameters. 
\end{enumerate}
Let us turn to the discussion of these points.

\subsection{Statistical procedure}

\subsubsection{Selecting the peaks}

Everybody would certainly agree that a peak pattern can be 
described as a price path which
first goes up, reaches a maximum, then goes down. However, real price
trajectories
present in addition many short term price fluctuations\,; this is
illustrated in Fig.1a,b,c. Fig.1c shows what the eye can
identify as a large isolated peak. However, its unambiguous definition
and the determination of the starting point of the raising pattern
depends on the time scale at which the data is coarse-grained.
If we subject the price series to a moving average procedure, local
fluctuations below this time scale will be smoothed out. The  width of the moving
average window thus determines the desired time scale. 
In the present paper, we deal exclusively with monthly
prices. We found that using a $41$-months window smooths out
adequately the local irregularities so as to get a quantitative
definition that parallels the intuitive and efficient pattern recognition
efficiency of the eye. We thus could eliminate the small scale structures
without affecting the overall shape of the large peaks. 
Varying the size of the window by about $20\%$ does not modify our results.

Let us summarize the procedure by which peaks have been systematically
selected\,: 
\begin{enumerate}
\item we have performed a moving
average with binomial weighting coefficients and a window-width of
41 months. The resulting series will be denoted by
$ p_m(t) $.  
\item We then computed the first differences $d_m(t) $
of the previous series. All successive times $ t_i $ for which $ d_m
(t_i) $ is stricly positive were considered as belonging to the same
rising path\,; decreasing paths were handled similarly. 
\item We then computed the amplitude, $ h= p_{max}/p_{min} $ of the peak
from the original series $ p(t) $ and we discarded the peaks for
which $ h $  was less than some critical value $ h_c $. In the
following $ h_c $ will be given the two values $ 1.5 $ and $
2.0 $. 
\end{enumerate}

This procedure is summarized in Fig.2 which provides the outcome of the
peak identification algorithm on a specific time series. The peaks and
troughs have been selected with
the threshold $ h_c = 2 $ and are indicated by heavy
dots. In addition to peaks and troughs, one could also consider
isolated rising or falling paths\,; these have been analysed in \cite{Roehner1989b}.
In the present paper, we restrict ourselves
to peaks and troughs.

\subsubsection{Definition of the variables describing the shape of the
peaks and troughs}

Two sets of variables will be considered in order to describe
the shape of the peaks. The first set summarizes the magnitude 
and symmetry of the peaks, the second set of parameters defines
their shape more precisely. 

In the first set, we consider five parameters defined as
follows\,; we shall here state the definitions for the case of
peaks but they can be easily deduced for the case
of troughs. 
\begin{enumerate}    
\item The amplitude $ h= p_{max}/p_{min} $ that has already been defined. 
\item The total duration $ d $ of the peak. 
\item The ratio $ r_p $ between the two prices at the beginning
and at the end of the peak: $ r_p = $ initial price/final price. 
\item The ratio $ r_d $ between the duration of the rising path
and the duration of the falling path: $ r_d =$ duration of the
path before the turning point/duration of the path after the
turning point.
\item The ratio $ r_s $ between the slopes of the rising path
and falling path. $ r_s = $ slope of the left hand section of the
peak/slope of the right hand section.
\end{enumerate} 

\vskip 0.5cm
The parameters $r_p, r_d$ and $r_s$ quantify the degree of (as)symmetry
between the rising and falling parts of a peak. The introduction of the
parameter $r_s$ is more specifically motivated by the well-known
result that Gaussian linear processes cannot rise to their maxima
and fall away at different rates \cite{Brockwell}.

All parameters should be invariant under rescaling, i.e. 
when all prices are
multiplied by a common factor. Otherwise those parameters would be
affected by the currency in which prices are expressed. 
The above parameters obviously satisfy this requirement. 
The Figures 1c, 4 and 6b  show that even the logarithms of the
prices exhibit a marked concavity. The goal of the second set of parameters
is to provide a measure of this concavity. For this purpose, we
propose the following representation\,:
\be
p(t) = A\exp \left[ -\hbox{sgn}(\tau )\left| {  t-t_0
\over \tau }\right| ^{\alpha} \right]  ~,
\label{eqe}
\ee
where $ t_0 $ denotes the turning point of the peak. 
\begin{itemize}
\item If $ \alpha $
is equal to $ 1 $, one retrieves an exponential growth up to the turning
point followed by an exponential decay. $x = \ln(p)$ is thus linear by part
with a tent-like structure.
\item If
$ \alpha <1 $ and $ \tau > 0 $, the function describes a
sharp peak of the kind represented in Fig.1c. 
\item If $ \alpha >1 $ and $ \tau <0 $, the function
describes a flat trough. 
\item If $ \alpha >1 $ and $ \tau >0 $, the
function describes a ``flat peak'' of a kind that will be seen to
exist in the real estate market
\item If $ \alpha <1 $ and $ \tau <0 $, the function describes a
sharp trough, a rare but not altogether inexisting phenomenon. 
\end{itemize}

These cases are summarized in Fig.3a. The parametrization (\ref{eqe})
is parsimonious and intuitive. The coefficient $A$ is just a scaling factor
that depends on the currency used for expressing the price. $t_0$ is the date
of the peak/trough, $|\tau|$ its duration and $\alpha$ quantifies the 
abruptness of the peak/trough. Close to $t_0$ (i.e. for 
$|t-t_0| \ll \tau$), the expression (\ref{eqe})
can be expanded in
\be
p(t)/A = 1  -\hbox{sgn}(\tau ) \left| { t-t_0
\over \tau }\right| ^{\alpha}  ~,
\label{eqedgq}
\ee
showing a power law behavior.

\subsection{Empirical results}

Table 1a lists the series for which we performed a statistical
analysis. 

\subsubsection{Peaks and troughs}

Table 1b summarizes the statistical findings for the first set of
parameters. The statistical results cannot rule out the null hypothesis
that the peaks are symmetric with respect to a vertical line 
drawn through the turning point. In
other words, after a speculative bubble, the
prices come back on average approximately to their initial level. 
Fig.3b provides a typical fit (to the peak price series in Munich from 1815 to
1820) using the parametrization (\ref{eqe}).

Table 1c summarizes our results for the values of 
the parameters $\alpha$ and $|\tau|$ that best fit the peaks and troughs 
listed in table 1c. We note the remarkably small dispersion of the
determined values of $\alpha$ around
\be
\alpha_{peak} = 0.63 \pm 0.03~.
\label{alpahdj}
\ee
Only the Vienna market is somewhat out of range with a larger $\alpha = 0.8$
and thus milder peak structure.

It is also noteworthy that no detectable variation of $\alpha$ or
$\tau$ can be detected in
the course of time. This becomes even more apparent if we list the
results for individual peaks (Table 1d).  Neither is there any 
correlation between $\alpha$ and the amplitude of the peak $h$.
In sum, the average values of $\alpha $ and $\tau $ for peaks
versus troughs are consistent with a SP-FT
pattern. 

\subsubsection{The turning point}

In the parametrization (\ref{eqe}), the turning point $t_0$ is a
singular point due to the absolute value (i.e. the expression of
$p(t)$ given by (\ref{eqe}) is not differentiable at $t_0$).
Is this singularity solely a result of this choice of
parametrization or 
does it correspond to some real economic features? We believe
that the latter holds and that the singularity reflects a genuine
cooperative behavior of the market not unsimilar to those 
studied in critical phenomena in Physics.

Indeed, during a major price peak, the spatial correlation length of
the markets, quantifying the correlations accross markets in different 
geographical areas,  tends to become very large. In ref.\cite{Roehner1989a},
it has been shown that during the price boost of 1816-1819 the
spatial correlation length jumped to a level about 20 times above its
``normal'' level. This could be interpreted in two different ways.
First one could think that it is the occurrence of an exogenous
perturbation that forced prices upward on all markets, thus producing
almost simultaneous price jumps throughout the country. Alternatively
one may interpret the jump in the correlation length as
reflecting a genuine endogenous increase of the strength
and range of the interactions between different markets.
To distinguish between the two interpretations,  
the correlation length of meteorological factors was calculated \cite{Roehner1989b}
and it was found that there is no
significant increase during or even
before the occurrence of the peaks. In further support to the second
interpretation of an endogenous cause, we note that it has been
 emphasized by many
historians \cite{Meuvret} that wheat trade expanded in
periods of high prices both in volume and in range. Locally,
many self-appointed retailers emerged\,; regionally,
traders were combing the countryside in order to buy every available
bushel of wheat for the supply of the cities (even at the
expense of the countryside supply)\,; internationally,
 the government often encouraged (and even
subsidized) the importation of additional quantities of wheat from
distant markets. 

\section{Twentieth century commodity markets}

\subsection{An overview}

Whereas in the previous section, it was possible to carry out a
systematic analysis with a relatively large data set,
the situation is much less favorable for the twentieth century.
For most commodities, there have been only a few major peaks during
this century. What makes things even worse is that
during the first half of the century, the world economy has been
disrupted by three major perturbations\,: World War I, the Great
Depression and World War II. For this reason, we restrict our
attention to the second half of the century. Our objective in this
section is to show by a few examples that the SP-FT
 pattern also applies to some twentieth commodity markets. A
more systematic investigation will be left to a subsequent paper.

In order to give an overview of twentieth century commodity
markets, we have listed in table 2 some of their main
characteristics. We note that the number of peaks in the interval 
considered is too small to give robust estimates of $\alpha$ and $\tau$
and the results are thus less reliable that for table 1c. In
order to minimize this difficulty, $\alpha$ and $\tau$ have been computed
only for the largest peaks with amplitudes larger than $4$. 
It should be emphasized that such data are taken
from various sources scattered throughout the literature\,; this is a
field for which a comprehensive handbook of empirical data would be
very valuable. 

There is a marked contrast between high- and low-volatility
commodities (recall that the volatility is the
economic term for the coefficient of variation, i.e. the ratio
of the standard deviation to the mean). The most volatile
commodity is sugar, whereas bananas are the least volatile.
Between those two products, there is a ratio of $3.4$ in terms of
coefficients of variation, and of $5.1$ in terms of the
amplitude of the largest peak. In the following, we examine two
cases\,: sugar and the precious metals, gold and silver. 

\subsection{Sugar}

Fig.4a depicts the shape of the three major peaks that occurred for
sugar prices in the interval 1950-1988. We get the following estimates
for the parameters $\alpha$ and $\tau$\,:

1962-1965: \quad $ \alpha =0.97, \quad \tau = 11.3 $

1972-1976: \quad $ \alpha =0.63, \quad \tau = \ 7.3 $

1978-1981: \quad $ \alpha =1.03, \quad \tau = 9.5$

Two of the values of $ \alpha $ are close to $1$, a
situation which was fairly rare for wheat markets in the previous century. 
We also note that $\tau$ is a factor of two or three smaller than
before.

\subsection{Gold and silver}

The case of the silver and gold bubbles which culminated in January 1980
is of special interest for at least three reasons.
\begin{enumerate}
\item These peaks were of
colossal dimensions both in breadth and in height\,: the duration was of
 the order of 20 years and the prices were multiplied by a factor of the 
order of 20\,; if the  amplitude of the peak shown in Fig.4b appears to be 
smaller, this is because it is based on annual average prices\,; in terms of
 weekly prices, the summit of the  peak is as high as 5000 cents/ounce. 
 \item Here, in contrast to many other cases, the phenomenon which triggered
 speculation is fairly well identified\,; it
was the fear of inflation which led many oil magnates to invest in precious
metals. This is quantitatively illustrated in Fig.4b\,; qualitatively it is well 
described in a fascinating book by S. Fay \cite{Fay}.
\item  Finally, among all
commodities, silver and gold have particularly low (relative)
 storage/transportation costs. This therefore provides strong evidence against
any theory of speculative bubbles which would solely rely on the impact
of storage/transportation costs. 
\end{enumerate}

\subsection{Price bubbles in other sectors}

In this paragraph, let us attempt a generalization. Indeed, it may be
illuminating to look at speculative phenomena for which 
``experimental'' conditions are somewhat different. This can help
us to separate the crucial variables from those which are specific
to a given environment, a prerequisite for a future theory of
collective speculative behaviour. 

\subsubsection{Real estate prices}

We consider in Fig.5 the case of real estate prices, a ``product''
characterized by a very small price elasticity of supply
especially in city centers, by long transaction times, and by 
fairly low storage costs. As a result, the real estate  price
bubbles turn out to follow a flat peak-flat trough pattern.
Indeed, the very small price elasticity of supply is probably
at the origin of a limited amplification of the speculative bubble,
therefore preventing an acceleration of the price.
There are counter examples, for instance in the speculative bubble of
Florida real estate market in the twenties (preceeding by a year the Oct. 1929 
stock market crash), in which the elasticity of supply was large because
new lands were plentiful as
previously considered useless lands were made avalaible to buyers \cite{Galbraith}.

\subsubsection{Bubbles in collector's items}

Price bubbles in collector's
items such as paintings, collectible autos or rare stamps often
parallel bubbles in the stock markets or in real estate markets\,;
here we shall restrict ourselves to illustrating this assertion by a
few examples.

\begin{itemize}
\item Table 3 shows the parallel evolution in the price of paintings
as computed by Buelens and Ginsburgh \cite{Buelens} and in the price of
stocks on the New York Stock Exchange. The amplitude of both peaks
is of the order of $3$ but the turning point in the price of paintings
occurred one year after the turning point for stocks. Incidentally,
it should be noted that the price boost of the late 1920s did not
simply reflect a general increase in the supply of money\,; in fact,
between 1920 and 1929 the total currency mass in the United States
increased by only 4.7\% \cite{Histo}.

\item In the late 1980s, there has been a speculative bubble in
collectible autos. Let us just mention one example\,: on the London
market, the price of a Ferrari 275 GTB4 built between 1966 and 1968
leapt from around $\$ \ 90'000$ in 1987 to $\$ \ 1$ million in 1989, a
multiplication by a factor of 11\,; by the end of 1991, it had fallen
back to $\$ \ 270'000$ (Herald Tribune, 28 May 1995). 

\item In Hong Kong, the prices of rare Chinese stamps have risen
steadily since 1970\,; yet, the trend accelerated at a dizzing pace
in the early 1990s. Let us mention two examples\,: the Chinese
1897 One Dollar stamp has risen from $\$ \ 1'570$ in 1970 to $\$ \
251'000$ in 1995, a multiplication by a factor 160\,; another stamp,
the Year of the Ram, had its price multiplied by a factor 430
between 1967 and 1995 (Herald Tribune 28 May 1995). 
\end{itemize}
Note in these examples the unexpectedly large
magnitude of the price jump. 

These are rather fragmentary indications\,; unfortunately the data
are too sparse and do not allow for an analysis of the shape of the
bubbles. Our aim in mentioning them was to emphasize that the
phenomenon of speculative bubbles extends beyond the spheres of
economics and finance. We think that 
this observation may be of interest in the perspective of
constructing a general theory of speculative behaviour.

\subsubsection{Stock market crashes}

Large stock market crashes are one of the most dramatic examples of speculative
bubbles culminating in peaks preceeded by an acceleration. 
It has been shown that the largest stock market crashes in this century
are outliers\,: they occur much more frequently that would predict the 
extrapolation of the historically determined distribution based on the more 
numerous smaller price variations, even when taking into account the significant
deviation from the Gaussian law \cite{crash8}. The apparent 
disappearance of the SP-FT structure in the present century
for the ``common'' peaks is paralleled by the very strong sharp peak 
pattern accompanying the largest crashes. This result suggests
 that large crashes result from 
amplification processes that have not been washed out by the large liquidity
of the modern markets. These amplification processes (of speculative nature) could be
 similar to those at the origin of the sharp peak structure on commodity prices 
 observed in the past centuries.

To pursue the analogy, we note that the exponents $\alpha$ reported in table
1c are remarkably close to those obtained for the largest crashes of this
century. When fitting the price $p(t)$ directly to a power law
(\ref{eqedgq}) (taking into account additional log-periodic 
corrections \cite{crash1}), we find
an exponent $\alpha \approx 0.3$, that is half the average value 
reported in table 1c. In contrast, if we fit the logarithm of the price to 
a power law (with again log-periodic corrections \cite{crash3}), we
find an exponent $\alpha \approx 0.6$ for the Oct. 1929, the Oct. 1987, the
May 1962 ``slow crash'' \cite{crash7},  the Hong-Kong Oct. 1997 crash
\cite{crashJohthesis} and the black monday of october 31, 1997 on the US 
equity market \cite{gainoption}. This value 
$\alpha \approx 0.6$ is remarkably similar to those found for the sharp
peaks of commodities in the previous centuries. Due
to increased liquidity and efficiency in modern markets, the sharp peak
patterns have essentially disappeared except for the most dramatic crashes
for which the precursory patterns develop over so long time scales (8 years
\cite{crash3}) that they have not yet been adequately arbitraged away 
by the market. The exponent $\alpha \approx 0.6$ might be the 
signature of the universality class of speculative/imitation behavior close
to the critical point (the turning point of the sharp peak).

\section{Toward a theory}

\subsection{Linear stochastic multiplicative models}

There is a large literature on the use of auto-regressive models to 
model economic times series. As already pointed out, the econometric nonlinear
autoregressive models with conditional heteroskedasticity 
(ARCH\,:  autoregressive-conditionally-heteroscedastic)  \cite{Engle} and
their generalizations \cite{Bollerslev} 
keep the volatility (standard deviation of price variations)
as the main descriptor and allow for the fact that the 
variance (or volatility)  is itself a stochastic variable.
In the simplest version of ARCH(1), we have the following
stochastic difference equation for the log-returns\,:
 \be
 R_{t+1} = \sqrt{b + a R_t^2} ~Z_t~,
 \label{arch}
 \ee
 where $Z_t$ is a gaussian random variable of zero mean and unit variance.
 This process (\ref{arch})
 describes a persistence and thus clustering of volatilities $R_t^2$. 
 Indeed, the factor
 $(b + a R_t^2)^{1 \over 2}$ ensures that the amplitude of the motion 
 $R_{t+1}$ is controlled by
 the past realization of the amplitude $R_t^2$. Now, calling 
 $X_t \equiv \langle R_t^2 \rangle$, 
 where the average is carried out over the realization of $Z_t$, 
 we see that (\ref{arch}) is equivalent to 
 \be
X_{t+1} = a X_t + b ~. \hspace{0.5in}     
\label{eq1}
\ee
We now allow the coefficient $a$ and $b$ to depend on
 time (independently from $Z_t$) leading to an equation with stochastic
 coefficient. This representation (\ref{eq1}) provides
 useful results under the following conditions. Let us assume that 
 $b$ is always positive and it may or may not fluctuate. Its presence ensures
 that $X$ does not shrink to zero asymptotically, even if the $a$'s are 
 less than one. We then imagine that the multiplicative factors $a$ are
 drawn from some distribution such that on average the rate of growth is
 negative, thus preventing the explosion of the process, but with intermittent
 realizations of $a$ larger than one. Intuitively, we can think that 
 the realization of $a$ reflects the meteorological factor, with $a<1$
 corresponding to favorable conditions while $a>1$ corresponds to bad
 weather with a negative impact on wheat production. $b$ can reflect a basic
 contribution of the price, like the minimum wage and price of production that
 may fluctuate but is strictly positive. If the meteorological conditions are
 always favorable, the realizations of $a$ are always less than one and $X$
 converges asymptotically to the fundamental price imposed by the price of 
 production.
 In contrast, in the presence of intermittent adverse factors, $a$ may become
 larger than one one for several time steps in a row, leading to a transient
 exponential growth. 
 
 This process (\ref{eq1}) has a long history. See \cite{physicaA} for a review with
 applications to population dynamics with external sources, epidemics,
 finance and insurance applications with relation to ARCH process, 
 immigration and investment portfolios, congestions on the internet,
 the statistical physics of directed polymers in random media, 
 auto-catalytic chemical reactions, etc. At first sight, 
 it seems that the linear model (\ref{eq1}) is so
simple that it does not deserve a careful theoretical investigation. However it turns
out that this is not the case\,: see, for example, a rather complicated
mathematical analysis of the problem in \cite{Kesten}. It turns out that model (\ref{eq1})
exhibits an unusual type of intermittency with a power law probability distribution
 of the variable $X_t$, for a large range of distributions for $a$ and 
$b$. The non-trivial properties of this
simple model (\ref{eq1}) come from the competition between the
multiplicative and additive terms \cite{Cont}.

The formal solution of (\ref{eq1}) for $t \ge 1$ can be obtained explicitly
\be
X_{t} = (\prod_{l=0}^{N-1} a(l)) X_0 + \sum_{l=0}^{N-1} b(l)
\prod_{m=l+1}^{N-1} a(m)~~,
\label{azjhg}
\ee
where, to deal with $l=N-1$, we define $\prod_{m=N}^{N-1} a(m) \equiv 1$. 
Because of the successive multiplicative operations on $a$ in the
iteration of $X_t$, it is clear that the behavior is locally exponentially
increasing or decreasing. The figure 6a shows a times series generated for $a$
and $b$ uniformly distributed in the intervals
$0.48 \leq a \leq 1.48$ and $0 \leq b \leq 1$. In this case $\langle \log a \rangle
= -0.06747$ and $\langle a \rangle = 0.98$.
The figure 6b shows the same realization with a logarithmic
scale in $X$, clearly demonstrating the tent-like peak structure. Thus, we 
conclude that stochastic multiplicative processes account for the intermittent
production of sharp peaks, but the upward concavity quantified by
(\ref{eqe},\ref{eqedgq}) with $\alpha = 0.6$ is not captured. In contrast to 
our empirical finding, this class of model predicts an exponent $\alpha = 1$.
This result is very instructive\,: stochastic multiplicative Kesten processes
account for a lot of observations, such as distributions of increments with fat
(power law) tails, self-affinity of the time series, multifractality, volatility
bursts, but they are not capable of representing the acceleration preceeding 
large peaks that are characterized by an exponent $\alpha$ significantly less than one.

\subsection{Nonlinear models}

We propose a nonlinear amplification phenomenon. To illustrate the idea, 
consider as a toy model the following
equation describing the price increments $\delta x$
\be 
{d \delta x \over d t} = \kappa (\delta x)^{\beta}~,
\label{mfg}
\ee
with $\beta > 1$, whose solution reads
\be 
\delta x(t) \sim |t_0 - t|^{1 \over \beta - 1}~.
\ee
Its time integral provides
\be 
x(t) \sim |t_0 - t|^{\alpha}
\ee
with 
\be
\alpha = {\beta - 2 \over \beta - 1}~,
\ee
which recovers (\ref{eqedgq}).
The observed value $\alpha \approx 0.6$ translates into $\beta \approx 3.5 $.
The intuitive interpretation of a value $\beta > 1$ is to represent multi-body
interactions between several players. The expression (\ref{mfg}) is then
similar to a ``mean-field'' equation representing the average behavior of
a representative agent interacting with an effective number $\beta$ of other
agents. This interaction, embodying the processes of imitation and herding, is
responsible for the observed acceleration preceeding large peaks.

While suggestive, this model remains very schematic and would need to be 
developed to account for the heterogeneity of the agents and the stochastic 
factors entering the market. This is left for a future work.

\section{Conclusion}

We have identified empirically a characteristic sharp peak-flat trough pattern
in a large set of commodity prices. These patterns provide a demonstration that 
those markets, that exhibit them, have not yet reached a fully efficient regime and
these patterns constitute recognizable signature of impending correlated price
series. We have shown that similar behaviors occur in a large variety of markets, 
all the more so, the less liquid is the market. Using simple models, we have
shown that nonlinear amplification processes must be invoked to account for the
observed acceleration. Mechanistically, the nonlinear behavior embodies multi-body
interactions leading to imitation and herding.

\vskip 1cm
\pagebreak

\pagebreak

Figure captions\,:
\vskip 1cm
Figure 1a\,: Monthly wheat prices in Toulouse (France) from 1500 to 1550. 
[not available in PostScript format] 

Figure 1b\,: Monthly wheat prices in Toulouse (France) from 1800 to 1850. 
[not available in PostScript format] 

Figure 1c\,: Weakly wheat prices in Munich during the 1815-1819 peak. 
[not available in PostScript format] 

\vskip 1cm
Figure 2\,: Peaks and troughs of amplitude over $100\%$ for wheat prices in Munich.
[not available in PostScript format] 

\vskip 1cm
Figure 3a\,: Peak and trough patterns.
[not available in PostScript format] 

Figure 3b\,: Least square fit of the generalized exponential (3) to the price
peak in Munich (1815-1819).
[not available in PostScript format] 

\vskip 1cm
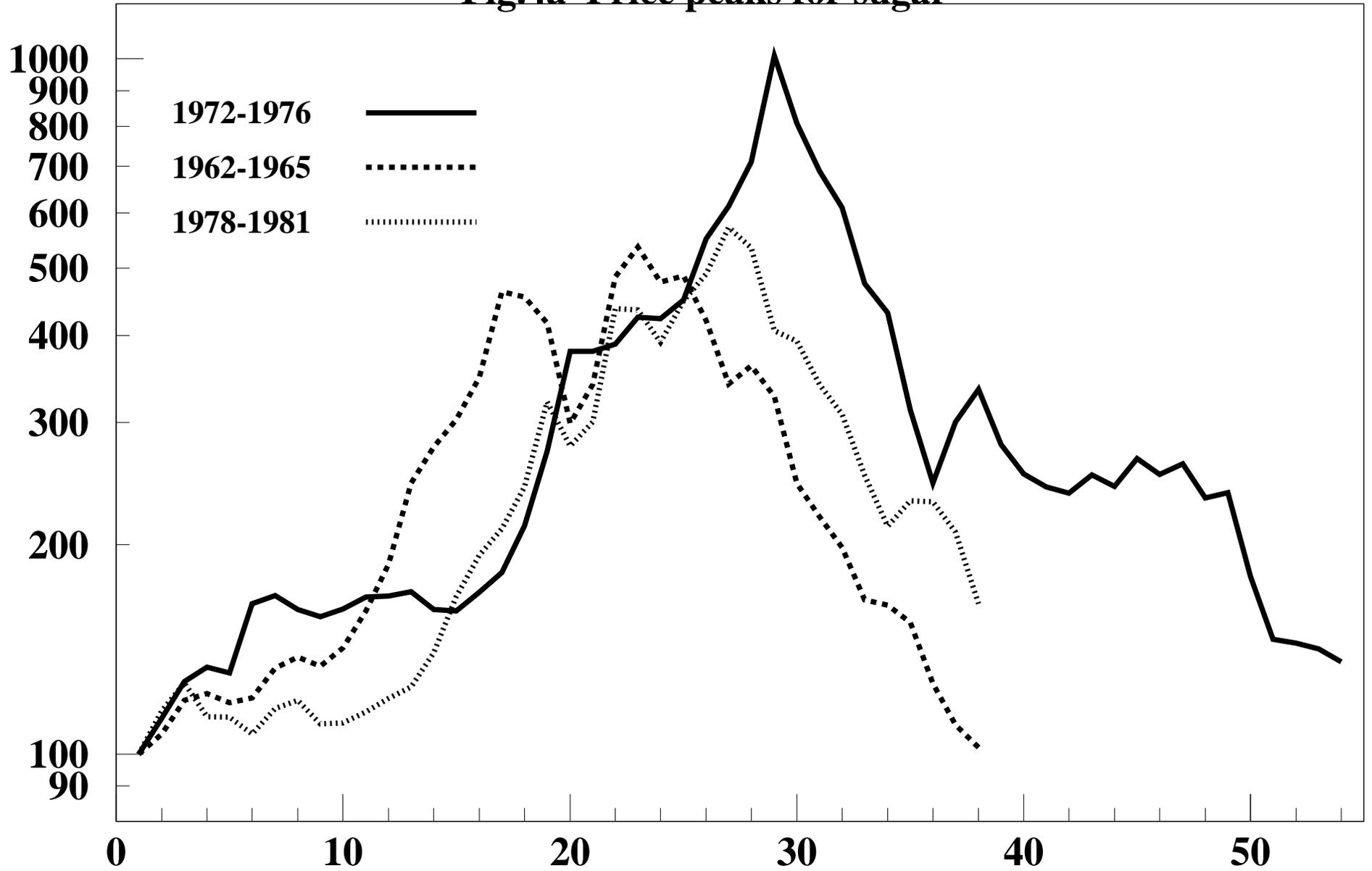
Figure 4a\,: Price peaks for sugar.

Figure 4b\,: Annual prices of silver in New York from 1950 to 1990 and inflation rates.

\vskip 1cm
Figure 5\,: Real estate price bubbles.
[not available in PostScript format] 

Figure 6a\,: Times series $X(t)$ generated for by eq.(\ref{eq1}) for $a$
and $b$ uniformly distributed in the intervals
$0.48 \leq a \leq 1.48$ and $0 \leq b \leq 1$. In this case $\langle \log a \rangle
= -0.06747$ and $\langle a \rangle = 0.98$. For these parameters,
$\langle X_t \rangle = {\langle b \rangle \over 1 - \langle a \rangle}
= 25$. Most of the time, $X_t$ is significantly less than its average,
while rare intermittent bursts propel it to very large values.

Figure 6b\,: Same as figure 6b but with a logarithmic scale for $X(t)$ 
showing a tent-like  structure of the peaks.


\begin{thebibliography}{99}

\bibitem{Arneodo} A. Arneodo, J.P. Bouchaud, R. Cont, J.F. Muzy,
M. Potters and D. Sornette,  Comment on ''Turbulent cascades in foreign exchange markets'', 
preprint cond-mat/9607120 at http://xxx.lanl.gov.

\bibitem{cascade} A. Arn\'eodo, J.-F. Muzy and D. Sornette,
``Direct'' cascade in the stock market, 
European Physical Journal B, in press (http://xxx.lanl.gov/abs/cond-mat/9708012)

\bibitem{Bachelier} M.L. Bachelier, {\em Th\'eorie de la Sp\'eculation}
(Gauthier-Villars, Paris, 1900).

\bibitem{Baulant} M. Baulant and J. Meuvret, Prix des c\'er\'eales extraits de la
Mercuriale de Paris (SEVPEN, Paris, 1960).

\bibitem{Biollay} L. Biollay, Le pacte de famine (Guillaumin. Paris, 1885).

\bibitem{Bollerslev} T. Bollerslev, R.Y. Chous and K.F. Kroner, 
J. Econometrics {\bf 52}, 5 (1992).

\bibitem{Beitrage} Beitr\"age zur Statistik Mecklenburgs 7, 3-4 (1873).

\bibitem{Bouchaud} J.-P. Bouchaud and M. Potters, Th\'eorie des risques
financiers (Al\'ea Saclay, Saclay, 1997).

\bibitem{Brockwell} P.J. Brockwell and R.A. Davis, Time series: theory and model
(Springer Verlag, 1987).

\bibitem{Buelens} N. Buelens and V. Ginsburgh, Revisiting Baumol's 'art as
floating crap game'. European Economic Review 37,1351-1371 (1993).

\bibitem{DeatonLaroque} A. Deaton and Laroque, On the behavior of commodity
prices, Review of Economic Studies 59,1-23 (1992).

\bibitem{Drame} S. Drame, C. Gonfalone, J.A. Miller and B. Roehner, Un si\`ecle
de commerce du bl\'e en France 1825-1913. Les fluctuations du champ des prix
(Economica, Paris, 1991).

\bibitem{Ebeling} D. Ebeling and F. Irsigler, 1976, Getreideumsatz, Getreide und
Brotpreise in K\"oln, Mitteilungen aus dem Stadtarchiv von K\"oln, Heft 65, K\"oln.

\bibitem{Farrel} M.L. Farrel, The Dow Jones averages 1885-1970 (Dow Jones Books, 
Princeton, 1972).

\bibitem{Fay} S. Fay, The great silver bubble, (London : Hodder and Stoughton,
1982).

\bibitem{Freche} G. Fr\^eche, Les prix de grains, des vins et des l\'egumes \`a
Toulouse (1486-1868), (Presses Universitaires de France, Paris, 1967).

\bibitem{Engle} R.F. Engle, Econometrica {\bf 50}, 987 (1982).

\bibitem{crash2} Feigenbaum, J.A., and P.G.O. Freund, Int. J. Mod. Phys. 10,
3737-3745 (1996).

\bibitem{crash4} J. A. Feigenbaum, P. G.O. Freund,
Discrete Scale Invariance and the ``Second Black Monday'',
cond-mat/9710324

\bibitem{Galbraith} J.K. Galbraith, 
The Great crash, 1929; with a new introduction
by the author (Boston : Houghton Mifflin, 1988).

\bibitem{Peinke} S. Ghashghaie, W. Breymann, J. Peinke, P. Talkner
and Y. Dodge, Nature {\bf 381}, 767 (1996).

\bibitem{crash6} S. Gluzman and V. I. Yukalov,
Renormalization Group Analysis of October Market Crashes,
cond-mat/9710336

\bibitem{Hissenhoven} P. van Hissenhoven, Le commerce international des grains
(Edited by the author, Brussels, 1923).

\bibitem{Histo} Historical Statistics of the United States 1975: U.S. Department
of Commerce. Washington.  

\bibitem{Huang} K.S. Huang, A complete system of U.S. demand for food (United
States Department of Agriculture, Technical Bulletin No 1821, 1993).

\bibitem{crashJohthesis} A. Johansen, Discrete scale invariance and other
cooperative phenomena in spatially extended systems with threshold dynamics, 
PhD thesis, CATS, Niels Bohr Institute, University of Copenhagen, Jan. 1998.

\bibitem{crash7} A. Johansen, O. Ledoit and D. Sornette,
Crashes as Critical Points, presented at the Annual Meeting of THE WESTERN
FINANCE ASSOCIATION,
June 17-20, 1998, Monterey, California, http://www.gsm.cornell.edu/wfa

\bibitem{crash8} A. Johansen and D. Sornette,
European Physical Journal B 1, 141-143 (1998).

\bibitem{Jump} P. Jorion, Review of Financial Studies {\bf 1}, 427-445 (1988).

\bibitem{Kesten} H. Kesten, Acta Math. 131, 207-248 (1973).

\bibitem{Labys1} W.C. Labys, Dynamic commodity models: specification,
estimation and simulation (Lexington Books. Lexington (Mass.)., 1973).

\bibitem{Labys2} W.C. Labys and P. Pollak, Commodity models for
forecasting and policy analysis. (Croomhelm Publishing. London, 1984).

\bibitem{Maillard} J.C. Maillard, Le march\'e international de la banane. Etude
g\'eographique d'un ``syst\`eme commercial'' (Presses Universitaires de Bordeaux,
Bordeaux, 1991).

\bibitem{Mandelbrot} B.B. Mandelbrot,
{\em Fractals and scaling in finance\,: discontinuity, concentration, risk}
(Springer, New York, 1997) and references therein.

\bibitem{Monthly1985} Monthly commodity price bulletin. 1960-1984 Supplement,
United Nations Conference on Trade and Development (UNCTAD), New York (1985).

\bibitem{Monthly1988} Monthly commodity price bulletin. 1970-1989 Supplement,
United Nations Conference on Trade and Development (UNCTAD), New York (1990).

\bibitem{Stanley} R. Mantegna and  H.E. Stanley,  Nature {\bf 376}, 46
(1995).

\bibitem{McNicol} D.L. McNicol, Commodity agreements and price stabilization 
(Lexington Books, Toronto, 1978).

\bibitem{Meuvret} J. Meuvret, Le commerce des grains et des farines 
Paris et les marchands parisiens  l'\'epoque de Louis XIV. Revue
d'Histoire Moderne et Contemporaine 3,169-203 (1956); 
Etudes d'histoire \'economiques. Cahier des Annales No 32 (1971).

\bibitem{Modeste} V. Modeste, De la chert\'e des grains et des pr\'ejug\'es
populaires qui d\'eterminent des violences dans les temps de
disette (Guillaumin. Paris, 1862).

\bibitem{Olsen} U.A. Muller, M.M. Dacorogna, R.D. Dav\'e, O.V. Pictet, R.B. Olsen
and J.R. Ward, UAM1993-08-16, O\&A Research Group preprint, XXXIXth International
conference of the applied econometrics association (AEA), Real time econometrics -
submonthly time series, 14-15 Oct.1993, Luxembourg.

\bibitem{Persson} K.G. Persson, On corn, Turgot, and elasticities: the case
of deregulation of grain markets in mid-eighteenth century France.
Scandinavian Economic History Review 1,37-50 (1993).

\bibitem{Pribram} A.F. Pribram, Materialien zur Geschichte der Preise und
L\"ohne in \"Osterreich (Carl Ueberreuters Verlag, Vienna, 1938).

\bibitem{Elliott} A.J. Prost and R. Prechter, ''Elliott waves principle'',
New Classic Library, 1985; T. B\'echu and E. Bertrand, ''L'analyse technique'',
Collection Gestion, Economica, 1992.

\bibitem{Radetzki} M. Radetski, A guide to primary commodities in the world economy
(Basil Blackwell, Oxford, 1990).

\bibitem{Roehner1989b} B.M. Roehner, Peaks and troughs in wheat price series.
Preprint LPTHE (November), unpublished (1989b).

\bibitem{Roehner1989a} B.M. Roehner, The decrease of price correlation with
distance and the concept of correlation length. Environment and
Planning A 21,289-298 (1989a).

\bibitem{Samuelson} P.A. Samuelson, {\em Collected Scientific Papers} 
(M.I.T. Press, Cambridge, MA, 1972).

\bibitem{Schultz} H. Schultz, The theory and measurement of demand (University
of Chicago Press, Chicago, 1938).

\bibitem{Seuffert} G.K.L. Seuffert, Statistik des Getreide und Viktualien Handels im 
K\"onigreiche Bayern, Weisz. Munich (1957).

\bibitem{physicaA} D. Sornette, Physica A 250, 295 (1998).

\bibitem{crash1} D. Sornette, A. Johansen and J.-P. Bouchaud, J.Phys.I
France 6, 167-175 (1996).

\bibitem{crash3} D. Sornette and A. Johansen, Physica A 245, 411 (1997).

\bibitem{Cont} D. Sornette and R. Cont,
J. Phys. I France 7, 431 (1997).

\bibitem{Stommel} H. Stommel and E. Stommel, The year without summer, Scientific
American, 134-140, June 1979.

\bibitem{Usher} A.P. Usher, The history of the grain trade in France
(Harvard University Press. Cambridge (Mass.), 1913). 

\bibitem{crash5} N. Vandewalle, Ph. Boveroux, A. Minguet and M. Ausloos,
The krach of October 1987 seen
as a phase transition\,: amplitude and universality, preprint 97

\bibitem{Weimar} F.H. Weymar, The dynamics of the world cocoa market
(MIT Press. Cambridge (Mass.), 1988).

\bibitem{gainoption} Prediction of the stock market turmoil at the end of october 1997,
 based on an unpublished extension of the log-periodic theory,
 have been formally issued ex-ante on september 17, 1997,
to the French office for the  protection of
proprietary softwares and inventions under number registration 94781.
See also H. Dupuis, ``Un krach avant novembre'',
Tendances, 18. September 1997,
page 26, from the work of N. Vandewalle, A.Minguet, P.Boveroux, and M. Ausloos
using the same type of log-periodic signals.


\end{thebibliography}
\end{document}